# The Evolution of the Bell Notion of *Beable*: from Bohr to Primitive Ontology

Federico Laudisa

Department of Humanities and Philosophy, University of Trento
Via T. Gar 14, 38122, Trento, Italy

**Abstract**

John S. Bell introduced the notion of *beable*, as opposed to the standard notion of *observable*, in order to emphasize the need for an unambiguous formulation of quantum mechanics. In the paper I show that Bell formulated in fact *two* different theories of beables. The first is somehow reminiscent of the Bohr views on quantum mechanics but, at the same time, is curiously adopted by Bell as a critical tool *against* the Copenhagen interpretation, whereas the second, more mature formulation was among the sources of inspiration of the so-called *Primitive Ontology* (PO) approach to quantum mechanics, an approach inspired to scientific realism. In the first part of the paper it is argued that, contrary to the Bell wishes, the first formulation of the theory fails to be an effective recipe for addressing the ambiguity underlying the standard formulation of quantum mechanics, whereas it is only the second formulation that successfully paves the way to the PO approach. In the second part, I consider how the distinction between the two formulations of the Bell theory of beables fares *vis-a-vis* the complex relationship between the theory of beables and the details of the PO approach.

# 1. Introduction

John Stewart Bell is unanimously recognized as one of the leading figures, if not *the* leading figure, of the foundational debate on quantum mechanics (QM) since the second half of the twentieth-century, a fact somehow posthumously acknowledged in the 2022 decision by the Nobel Committee to award physicists who designed and realized ingenious experimental tests of the Bell inequalities. He is also acknowledged as a fierce and relentless enemy of Copenhagenish approaches to QM: as is well known, his critical attitude toward any purely operational and instrumental understanding of quantum principles led him to encourage alternative views, ranging from Bohmian mechanics (starting from Bell 1966, his first work concerning the hidden variables' issue) to (idiosyncratic) forms of the Everett interpretation (Bell 1976), up to an explicit support to the so-called dynamical reduction model, or GRW version, of QM in the latest part of his career (1987, 1989, 1990). One of the most provocative proposals on the Bell part has been the introduction of the notion of *beable*, a term first introduced in 1973 with the specific aim of addressing what Bell took to be an intrinsic ambiguity in the quantum description of observation:

> This terminology, *be*-able as against *observ*able, is not designed to frighten with metaphysic those dedicated to realphysic. It is chosen rather to help in making explicit some notions already implicit in, and basic to ordinary quantum theory." (Bell 1975, in Bell 2004[2], p. 52).

The claim that a 'theory of' beables was needed, and its connection with the issue of locality, were the focus of the seminal papers of the Seventies in which Bell started to elaborate on the notion of beable. At the time the suggestion had not been taken too seriously, but the foundational role of beables has surfaced again in more recent times, when this notion turned out to be at the source of a true research program, the *primitive ontology* approach, in the area of the foundations and interpretations of quantum mechanics.

The primitive ontology approach emphasizes the need for a well-founded theory to specify in ontologically clear terms the kind of entity the theory itself is primarily supposed to account for[1]; within the framework of this approach, therefore, the term *beable* is ordinarily referred to as the expression of an attitude toward the foundations of quantum mechanics inspired to (some form of) scientific realism. When the notion of beable was proposed for the first time, however, no primitive ontology approach was around yet. It was clear that the proposal of a notion of beable by Bell was an expression of dissatisfaction toward the standard formulation of quantum mechanics, but it is far from transparent what the anti-instrumentalistic role, assigned by Bell to that notion, should have been exactly. The aim of the present work is to show that there are at least two different readings of the notion of beable in the development of Bell's foundational analyses, corresponding to an evolution in time of the interpretation that Bell provides for the notion itself. Initially – so I will argue – the concept of beable emerges as the consequence of a peculiar Bohrian-sounding view of the status and role of measurement in QM: I will show in section 2 that Bell, across several of his papers devoted to the foundations of QM, repeatedly and instrumentally exploits Bohr in different places, in order to support claims that in fact are meant to undermine the standard, Copenhagenish view of quantum mechanics. In these cases, Bell appears paradoxically to be using Bohr as a weapon *against* 'the Copenhagen spirit'[2]! Only later the Bell interpretation of the notion of beable evolves more explicitly into a second, more focused formulation: in section 3, I will emphasize that it is this new formulation that is apt to intertwine with the locality/non-locality issue arising from the formulation of the 1964 Bell theorem. In retrospect, therefore, we can recognize in this *second* stage of the Bell formulation of the notion of beable one strong motivation for the primitive ontology approach to the foundations of quantum mechanics. It will be also emphasized that, in spite of some not completely satisfactory wordings by Bell, the adoption of the notion of beable that will surface in the locality/non-locality issue clearly does not commit Bell to assume any form of naive 'realism', especially with respect to the so-called 'local realism' that, according to some, would be the alleged target of the Bell theorem. Finally, in section 4, the relation between the second stage in the Bell development of the notion of beable and the further evolution of the primitive ontology approach will be assessed.

---

[1] In what follows, with special reference to section 4, I treat the terms *ontology* and *metaphysics* interchangeably.

## 2  The Early History of Beables: Bell and Bohr

The first occurence of the term *beable* can be found in a short, programmatic Bell paper entitled "Subject and object" and published in 1973 (Bell 2004[2], pp. 40-44). First of all, the paper has a telling title. Bell decides to address the central role assigned to measurement in the standard formulation of quantum mechanics in terms of a distinction – that between 'subject' and 'object' – that has a *philosophical* tradition and flavor[3]. By pairing an *object* with a *measured system* and a *subject* with a *measurer*, Bell charges the standard formulation of quantum mechanics with a kind of subjectivism, according to which the theory is bound to retain a fundamental vagueness and ambiguity on where the boundary between subject and object is supposed to be located, no matter how good for practical use the theory is:

---

[2] It was Werner Heisenberg who already in 1930 used the expression *Kopenhagener Geist* (Copenhagen spirit) in the Preface to his book on the physical principles of quantum theory (Heisenberg 1930).
[3] According to a Bell biographer, the very title was a choice of the organizers of the conference in which the paper was first presented (Whitaker 2016, p. 290), but Bell employs the distinction with a conscious purpose.

> The subject-object distinction is indeed at the root of the unease that may people still feel in connection with quantum mechanics. […] In extremis the subject-object division can be put somewhere at the 'macroscopic' level, where the practical adequacy of classical notions makes the precise location quantitatively unimportant. But although quantum mechanics can account for these classical features of the macroscopic world as very (very) good approximations, it cannot do more than that. The snake cannot completely swallow itself by the tail. This awkward fact remains: the theory is only *approximately* unambiguous, only *approximately* self-consistent. (Bell 1973, in Bell 2004$^2$, pp. 40-41, emphasis in the original).

It is in expressing his hope in a less-and-less ambiguous formulation that Bell introduces for the first time the term *beable*:

> […] it should again become possible to say of a system not that such and such may be *observed* but that such and such *be* so. The theory would not be about '*observ*ables' but about '*be*ables'." (Bell 1973, in Bell 2004$^2$, p. 41).

Here in "Subject and object", Bell does not elaborate as precisely as one could wish a *theory* of beables, but we can interpret his wording as suggesting at least two conditions that such a theory should satisfy:

(i) although the use of the notion of beable cannot simply amount to make quantum mechanics a classical theory in any sense, a theory of beables should account for "an image of the everyday classical world", namely they should enable us – as middle-size natural systems – to recover our subjective experience;

(ii) at the same time, a theory of beables should justify the idea that beables somehow *ground*, or, even better, *constitute* observables: as Bell says with a sort of 'metaphysical' tone, "the idea that quantum mechanics is primarily about 'observables' is only tenable when such beables are taken for granted. Observables are *made out of* beables." (Bell 2004$^2$, p. 41).

Surprisingly, in order to support the plausibility of beables Bell appears to rely on a well-known passage of Niels Bohr, taken from the Bohr contribution to the 1949 celebrated volume *Albert Einstein Philosopher-Scientist*:

> […] it is decisive to recognize that, *however far the phenomena transcend the scope of classical physical explanation, the account of all evidence must be expressed in classical terms*." (Bohr 1949, p. 209, emphasis in the original).

Bell suggests not only that his notion of beable does justice to the Bohr plea for an account of evidence in classical terms, but also that – if formulated in terms of the beables' theory – such plea can be put to work in order to solve the above mentioned problem generated by the inherent ambiguity and approximation of standard quantum mechanics. The Bell suggestion is paradoxical, since it uses a major claim of the patriarch of the Copenhagen interpretation as a weapon *against* the Copenhagen interpretation itself: the theory of beables is introduced here clearly as an 'antidote' to the tendency to adopt an axiomatic formulation of quantum mechanics that relies essentially on an ill-defined (according to Bell) notion of measurement.

The use of the name of Bohr in the 1973 paper is not new to Bell, though. It occurs in the very first section of the first article devoted by Bell to the issue of hidden variables, namely the paper *On the problem of hidden variables in quantum mechanics*, written in 1963 but published in 1966. It is the path-breaking article in which Bell reviews the existing impossibility proofs for a hidden variable re-interpretation of quantum mechanics – from von Neumann 1932 to Jauch-Piron 1963, through the work of Gleason in 1957 – only to find them all wanting. As is well known, according to Bell all these proofs – no matter what the internal variants were – shared a common drawback, that of requiring assumptions that it was not reasonable to require from *any* possible, hypothetical hidden variable completion of quantum theory[4]. It is in the context of anticipating, in the first section, the core of the article that Bell exploits the name of Bohr, in order to support his claim and make the unreasonableness of the existing impossibility proofs even more apparent:

> It will be urged that these analyses [i.e. the above mentioned proofs] leave the real question untouched. In fact it will be seen that these demonstrations require from the hypothetical dispersion free states, not only that appropriate ensembles thereof should have all measurable properties of quantum mechanical states, *but certain other properties as well*. These additional demands appear reasonable when results of measurement are loosely identified with properties of isolated systems. They are seen to be quite unreasonable when one remembers with Bohr 'the impossibility of any sharp distinction between the behaviour of atomic objects and the interaction with the measuring instruments which serve to define the conditions under which the phenomena appear'. (Bell 1966, in Bell 2004[2], pp. 1–2, my emphasis).

The Bohr view, referred to by Bell, is that in a quantum measurement process a peculiar, non-classical form of non-separability emerges between object system and apparatus. In his 1966 paper Bell appears to exploit this Bohrian non-separability in support of his critical attitude toward the no-hidden variable theorems by von Neumann, Gleason and Jauch-Piron. In other words, Bell presents the Bohr view as an early instance of would have been called 'contextuality', suggesting at the same time that this should have long taught von Neumann, Gleason, Jauch and Piron that any serious hypothetical hidden-variable completion of quantum mechanics was bound to incorporate a form of context-dependence in the first place[5]. Given that Bohr was standardly conceived as the major representative of an approach to the foundations of quantum mechanics that could not be more alien to Bell in many respects, Abner Shimony has playfully described Bell's use of the Bohr claim:

---

[4] To a large extent, this is still the folklore view in the community of the philosophy and foundations of quantum mechanics. However, the issue of a correct interpretation of what Bell exactly argued against the von Neumann, Gleason and Jauch-Piron theorems, and to what extent the Bell criticisms were justified, is far from trivial and still debated. For the Bell arguments against von Neumann-Gleason and Jauch-Piron see, respectively, Acuna 2021, and Laudisa 2023.

[5] The meaning and role of contextuality in the Bohr philosophy of quantum mechanics is a relevant issue in the Bohrian scholarship: see for instance the essays by Jan Faye, Mauro Dorato and Dennis Dieks in Faye, Folse 2017.

Bell, *by a judo-like manoeuvre*, cited Bohr in order to vindicate a family of hidden variables theories in which the values of observables depend not only upon the state of the system but also upon the context." (Shimony 1984, in Shimony 1993, p. 121, my emphasis).

In the same vein, the name of Bohr emerges in the 1971 Bell paper *Introduction to the hidden-variable question*, where Bell first introduces the family of stochastic hidden variable theories. In discussing "the very essential role of apparatus" in the quantum-mechanical description of the measurement process, Bell argues that

> The result of the measurement does not actually tell us about some property previously possessed by the system, but about something which has come into being in the combination of system and apparatus. *Of course, the vital role of the complete physical set-up we learned long ago, especially from Bohr*." (Bell 1971, in Bell 2004[2], p. 35).

Bell returns to the same point in his later 1982 article "The impossible pilot wave". In recalling once again the lack of generality of the early no-hidden variable theorems, Bell writes about what he calls 'the Gleason-Jauch argument':

> For a given operator $P_1$ it is possible (when the dimension $N$ of the spin space exceeds 2) to find more than one set of other orthogonal projection operators to complete it:
> $$1 = P_1 + P_2 + P_3 \ldots$$
> $$= P_1 + P'_2 + P'_3 \ldots$$
> Where $P'_2 \ldots$ commute with $P_1$ and with one another, but not with $P_2 \ldots$. And the extra assumption is this: the result of 'measuring' is independent of which complementary set … or … is 'measured' at the same time. The de Broglie-Bohm picture does not respect this. […] In denying the Gleason-Jauch independence hypothesis, the de Broglie-Bohm picture illustrates rather the importance of the experimental set-up as a whole, *as insisted on by Bohr. The Gleason-Jauch axiom is a denial of Bohr's insight.* (Bell 1982b, in Bell 2004[2], p. 165).

We have evidence, then, that Bohr has a place in the Bell line of thought about the foundations of quantum mechanics already in the early Sixties, as a forerunner of the idea of contextuality.

But let us return to what we called the Bell theory of beables, as expressed by the conditions (i) and (ii). These conditions appear far from uncontroversial, when referred to the early, Bohrian characterization of beables by Bell. If condition (i) sounds milder, since it seems to require just compatibility with common sense, condition (ii) is more puzzling. What sort of 'constitution' property is supposed to be involved in the claim that observables are 'made out of' beables? What are beables supposed to be in order to 'make up' observables? And what is the exact relation of such intuition of 'constitution' with the Bohrian view of quantum measurements? The paradox – the use of Bohr *against* Copenhagen quantum mechanics – would do no harm as such, but the Bell strategy is dubious by a *conceptual* point of view. I wish to argue that the Bohrian requirement to express experimental evidence in 'classical' terms, in order for linguistic communications among scientists to be consistenly preserved, can hardly be put usefully to work to provide the unambiguous description of the quantum measurement process that Bell was searching for.

The extent to which the reference to Bohr may really play the role of dissolving the ambiguity deplored by Bell is a matter of dispute, since it concerns the status of an issue that is still debated in the reconstructions of the Bohr attitude toward quantum mechanics: the issue of whether, according to Bohr, quantum mechanics should be taken as universal – i.e. applicable to *all* physical systems, including measuring instruments – or not. The problem of the universality of quantum mechanics in principle emerged since the very origins of quantum theory, due to the increasing divergence from all preceding classical physics that was apparent in the experimental development of the theory already in the first decades of the twentieth century. In the early days of the debate on the foundations of quantum mechanics, it was far from clear what the relation between the classical and the quantum regimes was supposed to be, until the mathematical treatment of the theory in the 1932 von Neumann treatise allowed physicists to put the problem in a clearer light in terms of the notorious 'measurement problem', raising for the first time the universality issue for quantum mechanics. The von Neumann treatment, and the place occupied by this problem in his first formally rigorous formulation of quantum theory, already revealed how controversial the status of measurement in quantum mechanics would have been, to the extent that the very notion of measurement would turn out to be the *locus classicus* for emphasizing the lack of consensus on the interpretation of the theory: von Neumann explicitly confronts the implications of the assumption that – in the context of a measurement of a physical quantity on a quantum system $S$ with an apparatus $A$ – the laws of QM govern *both S and A*. This view has acquired with time the status of a commonplace: 'quantum fundamentalism' – this is how, for instance, Zinkernagel 2015 calls it – is the claim that "Everything in the universe (if not the universe as a whole) is fundamentally of a quantum nature and ultimately describable in quantum-mechanical terms." In Zinkernagel's words:

> In this formulation, quantum fundamentalism contains both an ontological and an epistemological thesis: that everything is of a quantum nature is an ontological claim, whereas the idea that everything can (at least in principle) be *described* in quantum terms is epistemological. The ontological component of quantum fundamentalism can also be expressed as the idea that we live in a quantum world. (Zinkernagel 2015, p. 41, emphasis in the original).

In fact Bohr never discussed explicitly the measurement problem in the von Neumann formal context. A wide consensus was established among most Bohr scholars, however, according to which his overall philosophical outlook legitimates a *non*-universalistic reading of quantum mechanics, mainly due to the special role attributed to classical categories in accounting for the experimental evidence in quantum measurements. For instance in a recent, qualified defense of this consensus, Zinkernagel 2015 refers to a 1938 paper in which Bohr argues that

> […] in each case some ultimate measuring instruments, like the scales and clocks which determine the frame of space-time coordination – on which, in the last resort, even the definitions of momentum and energy quantities rest – must always be described *entirely* on classical lines, and consequently kept outside the system subject to quantum mechanical treatment." (Bohr 1938, p. 104, emphasis in the original).

One can make sense of this argument, according to Zinkernagel, only under the assumption that quantum mechanics actually *fails* to be universal:

> A way to understand Bohr's requirement is that we need a reference frame to make sense of, say, the position of an electron (in order to establish with respect to what an electron has a position). And, by definition, a reference frame has a well-defined position and state of motion (momentum). Thus the reference frame is not subject to any Heisenberg uncertainty, and it is in this sense (and in this context) classical. This does not exclude that any given reference system could itself be treated quantum mechanically, but we would then need another – classically described – reference system e.g. to ascribe position (or uncertainty in position) to the former. (Zinkernagel 2015, p. 430)[6].

This view has been challenged. Already Landsman 2007, for instance, had argued that the Bohr texts would not justify an interpretation of his thought to the effect that there exists an independent natural realm of an intrinsic classical character. Let us consider the following passage, contained in a famous Bohr paper entitled "On the notions of causality and complementarity", published in 1948 on the philosophical journal *Dialectica*:

> The construction and the functioning of all apparatus like diaphragms and shutters, serving to define geometry and timing of the experimental arrangements, or photographic plates used for recording the localization of atomic objects, will depend on properties of materials which are themselves essentially determined by the quantum of action. (Bohr 1948, p. 145).

On the basis of texts like this, Landsman claims that the division system/apparatus, in which the former is described quantum-mechanically whereas the latter is described classically, has no *ontological* import:

> there is no doubt that both Bohr and Heisenberg believed in the fundamental and universal nature of quantum mechanics, and saw the classical description of the apparatus as a purely *epistemological* move, which expressed the fact that a given quantum system is being used as a measuring device" (Landsman 2007, p. 437, emphasis added).

In a recent contribution Dieks reinforces this challenge, defending an exclusive *epistemic* reading of the role of the classical notions in the Bohr view of the quantum measurement process, denying any *ontological* quantum non-universalism by Bohr (Dieks 2017).

This dispute on the ontological or epistemological flavor of the quantum/classical divide, however, leaves the ambiguity point that concerns us here untouched. We do not need to take a stance on whether the boundary between the classical and the quantum world concerns our knowledge or the ultimate structure of Nature to see that we are forced anyway, within the Heisenberg-Bohr Copenhagen framework, to acknowledge that, on one hand, we cannot but locate somewhere the infamous 'cut' between system and apparatus, and on the other hand there is no rigorous recipe even on a pragmatic level about where *exactly* we should put it. As Dieks himself remarks, in the very first section of the seminal complementarity paper published in 1928, Bohr emphasizes that

> The circumstance [...] that in interpreting observations use has always to be made of theoretical notions entails that for every particular case *it is a matter of convenience* at which point the concept of observation involving the quantum postulate with its inherent "irrationality" is brought in." (Bohr , p. 54, emphasis added).

---

[6] A more sustained defense of this view is contained in Zinkernagel 2016. A recent interpretation of quantum mechanics that reads Bohr in a non-universalistic fashion, in order to support its own account of the measurement process, is the most up-to-date version of the Bub information-theoretic interpretation (Bub 2018).

Wolfgang Pauli echoed the same point in a 1949 paper, entitled "The Philosophical Significance of the Idea of Complementarity":

> [...] modern physics generalizes the old placing in opposition of apprehending subject on one hand and object apprehended on the other to the idea of the cut between the observer or instrument of observation and the system observed. While the existence of such a cut is a necessary condition of human cognition, modern physics regards the position of the cut as to a certain extent arbitrary, and as the result of a choice partly determined by considerations of expediency, and therefore partly free. (Pauli 1950, p. 41, emphasis added).

As a consequence, the Bell judo-like manoeuvre in this case can hardly work, that is, the 'ambiguity' and 'approximation' of the standard formulation of quantum mechanics cannot be relieved by the use of the Bohrian framework. In particular, the Bohrian model of the quantum measurement may at most satisfy the Bell condition (i), namely, the 'functionalistic' recovery of subjective experience, but fails to satisfy unambiguously the 'constitutive' Bell condition (ii), since the concrete individuation of the relevant beables depends on *arbitrary* criteria: with the resources allowed by the Bohr framework, quantum observables simply *cannot* be 'made out' of beables.

## 3  From Bohr to Primitive Ontology: the New Life of Beables

In the first appearance of the notion of beable, the early Bell move – use Bohr *against* Copenhagen quantum mechanics – looks then rather unfortunate. But the role that we have analyzed in the previous section starts to be replaced in the subsequent development of the notion itself. For Bell returns to beables in a 1975 paper, whose title ("The theory of local beables") this time mentions explicitly the need for a *theory* of these 'objects', whatever they are meant to be. At first sight, the very opening of the paper is in line with the Bohrian attitude we have alluded to:

> This is a pretentious name for a theory which hardly exists otherwise, but which ought to exist. The name is deliberately modelled on 'the algebra of local observables'. This terminology, *be*-able as against *observ*able, is not designed to frighten with metaphysic those dedicated to realphysic. It is chosen rather to help in making explicit some notions already implicit in, and basic to ordinary quantum theory. For, in the words of Bohr, ' it is decisive to recognize that, however far the phenomena transcend the scope of classical physical explanation, the account of all evidence must be expressed in classical terms.' It is the ambition of the theory of local beables to bring these 'classical terms' into the equations, and not relegate them entirely to the surrounding talk. (Bell 1975, in Bell 2004$^2$, p. 52).

In clarifying what beables are supposed, or meant, to be, Bell refers again to macroscopic pieces of experimental settings in a broad sense – and this is, once again, entirely Bohrian in spirit – but, *this time*, he expresses explicitly the need for a clear *theory* of them, in terms of a more robust sense of physical reality:

> The beables must include the settings of switches and knobs on experimental equipments, the current in coils, and the readings of instruments. 'Observables' must be *made*, somehow, out of beables. The theory of local beables should contain, and give precise physical meaning to, the algebra of local observables. (Bell 1975, in Bell 2004$^2$, p. 52).

This appears to be a turning point in the Bell characterization of beables. Not only Bell refers to the difference in electromagnetism between 'physical' entities (like the electric and magnetic fields) and 'unphysical' entities (like potentials), in order to set up a distinction according to which beables should be clearly located on the 'physical' side. He also points here to what we have called above a condition of 'constitution', a more fundamental status that beables should be endowed with: it is this status that in principle justifies the observables being *made out* of beables. This conjunction of realism – beables *are out there* – and constitution – beables are what *make up* observables and all that gravitates around observation – characterizes the new Bell theory of beables, and his later paper "Beables for quantum field theory" (1984) testifies it:

> There is nothing in the mathematics to tell what is 'system' and what is 'apparatus', nothing to tell which natural processes have the special status of 'measurements'. Discretion and good taste, born of experience, allow us to use quantum theory with marvelous success, despite the ambiguity of the concepts named above in quotation marks. But it seems clear that in a serious fundamental formulation such concepts must be excluded. In particular we will exclude the notion of 'observable' in favour of that of '*be*able'. The beables of the theory are those elements which might correspond to elements of reality, to things which exist. Their existence does not depend on 'observation'. Indeed observation and observers must be made out of beables. (Bell 1984, in Bell 1987, p. 174).

That beables should correspond "to elements of reality, to things which exist" might still sound compatible with the Bell early, Bohrian-sounding formulation that we analyzed in the previous section[7], but clearly this is not the case with the claim that the existence of beables *does not depend*

---

[7] This is true also of what Bell says a page later when he claims: "Not all 'observables' can be given beable status, for they do not all have simultaneous eigenvalues, i.e. do not all commute. It is important to realize therefore that most of

*on 'observation'*: in Bohrian terms, on the contrary, it is exactly the reference to the context of observation that allows macroscopic pieces of experimental settings (namely, what Bell takes as beables in his early formulation) to be part of a scientifically meaningful experience[8].

In connection with this emphasis *both* on the 'reality' of beables and their 'constitutive' nature, Bell introduces for the first time a connection with an intuitive sense of *locality*, called here *local causality*[9]:

> We will be particularly concerned with *local* beables, those which (unlike the total energy) can be assigned to some bounded space-time region. […] It is in terms of local beables that we can hope to formulate some notion of local causality. (Bell 1975, in Bell 2004[2], p. 53, emphasis in the original).

It is *this* focus on locality – I argue – that determines a new twist for the formulation of a theory of beables, a formulation which starts to diverge from the Bohrian-sounding notion reviewed in the previous section and receives a more distinctive 'fundamental' status in somewhat ontological terms. Bell attempts to figure out a definition of local causality that can work also in an indeterministic setting, an attempt that leads him to introduce an expression like $\{A \mid \Lambda\}$, that stands for the probability of a particular value $A$, given particular values $\Lambda$ (Bell 1975, in Bell 2004[2], p. 54). An interesting point to note here is that, in introducing this expression, Bell employs the term 'beable' to denote a *value* (of a physical quantity), something very different from "settings of switches and knobs on experimental equipments", which was the original, Bohrian-sounding meaning attached to the term. On this new background Bell operates in a much more explicitly 'realistic' (and much less 'Bohrian') vein – a background in which it is perfectly sensible to conceive an observer-independent world whose unveiling is a major task for fundamental physics – and the new reading of beables in terms of values is immediately put to work in an EPR-kind of context:

> Let *A* be localized in a space-time region 1. Let *B* be a second beable localized in a second region 2 separated from 1 in a spacelike way. Now my intuitive notion of local causality is that events in 2 should not be 'causes' of events in 1, and viceversa. But this does not mean that the two sets of events should be uncorrelated, for they could have common causes in the overlap of their backward light cones. It is perfectly intelligible then that if $\Lambda$ in 1 does not contain a complete record of events in that overlap, it can be usefully supplemented by information from region 2. So in general it is expected that
> $$\{A \mid \Lambda, B\} \neq \{A \mid \Lambda\}$$
> However, in the particular case that $\Lambda$ contains already a *complete* specification of beables in the overlap of the two light cones, supplementary information from region 2 could reasonably be expected to be redundant. (Bell 1975, in Bell 2004[2], p. 54, emphasis in the original).

---

these 'observables' are entirely redundant. *What is essential is to be able to define the position of things, including the positions of instrument pointers or (the modern equivalent) of ink on computer output.*" (Bell 1984, in Bell 2004[2], p. 175).

[8] This second formulation of the Bell theory of beables has its last expression, in chronological terms, in the section 3 of his (amusing) paper "La nouvelle cuisine", published in 1990 (Bell 2004, pp. 232-248).

[9] As already remarked by others, this expression is likely to be misleading in suggesting that the influence at stake *should* have a direction, which in fact is not necessarily the case.

It is quite clear, then, that the above mentioned specification of beables makes sense in the Bell second, ontologically-loaded formulation of the notion of beable (and not in the old, Bohrian-sounding one). Moreover, this new formulation is immediately put to work in the investigation on whether, in the Bell language, quantum mechanics might be shown to be 'locally causal' if reformulated as a sub-theory of a 'more complete' theory:

> But could it not be that quantum mechanics is a fragment of a more complete theory, in which there are other ways of using the given beables, or in which there are additional beables – hitherto 'hidden' beables? And could it not be that this more complete theory has local causality? Quantum mechanical predictions would then apply not to given values of all the beables, but to some probability distribution over them, in which the beables recognized as relevant by quantum mechanics are held fixed. We will investigate this question, and answer it in the negative. (Bell 1975, in Bell 2004[2], p. 55).

When taken as representative of what later has become known as the locality/non-locality issue, however, the above quotation by Bell is not as clear as one – in retrospective – could wish. In fact, Bell appears here to present the program (to which, through a stochastic generalization of his 1964 theorem, he will answer in the negative) as a program that starts from envisaging a super-theory in which all 'additional' values are *already in place*, in particular all the values that are needed in order, so to say, to fill up the gaps left by quantum mechanics. This way of presenting the problem, though, has a serious drawback: it suggests that, in order to derive an inequality that turns out to be violated by quantum-mechanical probabilities, Bell needs to assume both locality *and* the pre-existence or definiteness of these 'additional' values. This would remarkably affect the interpretation of the Bell theorem and of the violation of the inequality entailed by that theorem, since under this interpretation this violation might be explained by a failure of locality *but also* by a failure of pre-existence or definiteness. In other terms, this is the problem of what the set of assumptions in the derivation of the Bell theorem exactly is, a problem that does not cease to be a matter of dispute[10].

We can find a further instance of the oscillation by Bell on this point: it is the Appendix of a Bell 1976 paper, entitled "Einstein-Podolsky-Rosen experiments" (Bell 1976, in Bell 2004[2], pp. 81-92), in which Bell summarizes a controversy with the well-known and influential historian of physics Max Jammer, who had expressed divergent views in his 1974 book *The Philosophy of Quantum Mechanics*. In the chapter 7 of the book, devoted to hidden variable theories, Jammer presents his view of the relation between the Einsteinian reservations on quantum mechanics and the hidden variable program, wondering to what extent the former might have been the main factor for the development of the latter:

> Although the Einstein-Podolsky-Rosen incompleteness argument was undoubtedly one of the major incentives for the modern development of hidden variable theories, it would be misleading to regard

---

[10] For a recent review cp. Lambare 2022.

Einstein, as some authors do, as a proponent or even as "the most profound advocate of hidden variables"[11]. True, Einstein was sympathetically inclined toward any efforts to explore alternatives, and as such the ideas of de Broglie and of Bohm, but he never endorsed any hidden variable theory. […] No doubt, Einstein's criticisms, and in particular his work with Podolsky and Rosen greatly contributed to the development of hidden variable theories, just as Mach's ideas contributed to the rise of Einstein relativity: but, as is not uncommon in the history of physics, the intellectual originator of a theory does not necessarily identify himself with its full-fledged development. (Jammer 1974, pp. 254-255)

In particular, Jammer points at the Bell 1964 paper as a major source for what he takes to be a misleading view, when in a footnote to the above quoted passage he refers to the opening lines of the Bell paper (already quoted in section 3), where the author writes that "[t]he paradox [i.e. the EPR argument] was advanced *as an argument that quantum mechanics should be supplemented by additional variables*." (Jammer 1974, p. 254, fn 3, my emphasis). In the Appendix to his 1976 paper Bell cites *with approval* the very same Shimony characterization of Einstein as "the most profound advocate of hidden variables" that Jammer had attacked, and replies to the Jammer footnote. Still in this footnote, Jammer had further claimed that some Einstein remarks, taken by Bell 1964 to support his view of hidden variable theories as *direct* consequence of the EPR argument, could not be read as "confessions of the belief in the necessity of hidden variables" (Jammer 1974, p. 254, fn 3). What was the Einstein remark?

> But on one supposition we should, in my opinion, absolutely hold fast: the real factual situation of the system $S_2$ is independent of what is done with the system $S_1$, which is spatially separated from the former. (Einstein 1959, p. 85)

In order to justify his use of this remark at the beginning of his 1964 paper, Bell writes that

> the object of this quotation was to recall Einstein's deep commitment *to realism and locality, the axioms of the EPR paper*." (Bell 1976, in Bell 2004[2], p. 89, my emphasis).

and it is important to recall that Bell quotes the Einstein remark in a footnote to a passage where he claims that "these additional variables were to restore to the theory *causality and locality*" (Bell 1964, in Bell 2004[2], p. 14, my emphasis), where *causality* clearly, although in a rather sloppy way, plays the role of what Bell calls *realism* in the 1976 Appendix. As is clear from this wording, then, Bell himself cannot avoid to run afoul of the very same mistake that *he* would have denounced himself a few years later, most clearly in his 1982 paper "Bertlmann's socks and the nature of reality":

> It is important to note that to the limited degree to which *determinism* plays a role in the EPR argument, it is *not assumed* but *inferred*. What is held sacred is the principle of 'local causality'—or 'no action at a distance'. [. . .] It is remarkably difficult to get this point across, that determinism is not a *presupposition* of the analysis." (Bell 1982a, in Bell 2004[2], p. 143, emphasis in the original).

---
[11] The reference is to Shimony 1971 (reprinted in Shimony 1993, p. 87).

And in a footnote few lines later he explicitly writes:

> "My own first paper on this subject [Bell refers here to his 1964 paper] starts with a summary of the EPR paper *from locality to deterministic hidden variables*. But the commentators have almost universally reported that it begins with deterministic hidden variables." (Bell 1982a, in Bell 2004[2], p. 157, emphasis in the original).

Remarkably, in saying in the 1976 Appendix that 'realism' is *also* an 'axiom' of the EPR paper, Bell is doing here exactly what "the commentators have almost universally reported"!

Now it is true that, as I attempted to show, these two passages from the two Bell papers from 1975 and 1976 seem to licence a 'local-realistic' reading by Bell himself. It is also true, however, that the Bell texts that seemingly justify an endorsement for local realism occupy in fact a measure-zero set – so to say – with respect to all other Bell pronouncements in favour of the view according to which, in the logical structure of the Bell theorem, the 'realism' condition is either derivative (in the strict anticorrelation case) or irrelevant (in the non-strict anticorrelation case): it is just locality that matters, namely the condition according to which, given two spacetime regions *A* and *B* that are space-like separated, events located in *A* cannot influence events located in *B* and viceversa. It is to be stressed that this is a very general condition of locality, and this generality must be taken into due account to address a basic objection, that might appear to threaten the consistency of such condition. According to this objection, the very act of assuming locality presupposes the assumption of objective, pre-existing properties for the systems located in the regions *A* and *B*: if I assume that there is no influence going on between *A* and *B* – so the objection goes – it means that I assume that at *A* and *B* there already *are* definite properties of physical systems, otherwise what is there that *can* be influenced when I discover that the Bell inequality is violated by quantum mechanics? The objection is not compelling, though. Clearly, in the context of a discussion of the physical framework suitable to the formulation of the Bell theorem, we need to assume that it makes sense to perform measurements, in order to obtain an unambiguous outcome at the end of the measurement procedure. In this sense, we cannot avoid to assume a minimal sense of 'stuff', existing out there in the natural world we are trying to investigate, an assumption without which the very enterprise of physics as a natural science would hardly make sense. We might, in this sense, use the term *event* and conceive locality as the assumption that there is no superluminal influence across distant 'events': that is, nothing that goes on at *A* (*B*) can affect what goes on at *B* (*A*). But this is far more general and minimal than assuming a strong notion of realism, namely something like "Every quantum observable actually has a definite value even before any attempt to measure it; the measurement reveals that value" or something like "The outcome of every experiment is pre-determined by some ("hidden") variable λ". It is *this* strong form of realism that need *not* be assumed in order to prove the Bell theorem but that in my view is often mistakenly assumed as an independent condition along with locality in some versions of the Bell theorem.

# 4 The Bell beables and the Primitive Ontology framework: a complex relationship

As a matter of fact, the Bell notion of beable was a major source of inspiration for the so-called primitive ontology (PO) approach, a global interpretive framework that was first proposed in 1992 by Detlef Dürr, Sheldon Goldstein and Nino Zanghì (DGZ). In their view, a firm foundation of quantum mechanics should methodologically presuppose as a preliminary step the unambiguous selection of the set of "the basic kinds of entities that are to be the building blocks of everything else" (DGZ 1992, p. ) and, second, the formulation of a (set of) law(s) governing the time evolution of the above mentioned building blocks. So the PO approach can then be conceived as a two-step approach in which, first, an ontology is fixed and, second, a law is provided for how the basic constituents of the ontology change over time. In a more up-to-date presentation:

> According to the primitive ontology (PO) approach, all fundamental physical theories have a common structure, which provides a general explanatory schema with which the theory accounts for what the world is like. According to this approach, any satisfactory fundamental physical theory, if taken from a realist point of view, contains a metaphysical hypothesis about what constitutes physical objects, the PO, which lives in three-dimensional space or space-time and constitutes the building blocks of everything else. In the formalism of the theory, the variables representing the PO are called the primitive variables. In addition, there are other variables necessary to implement the dynamics for the primitive variables: these non-primitive variables could be interpreted as law-like in character. Once the primitive and the non-primitive variables are specified, one can construct an explanatory schema based on the one that is already in use in the classical framework. This allows determining, at least in principle, all the macroscopic properties of familiar physical objects in terms of the PO. This structure holds for classical as well as for quantum theories. (Allori 2015, pp. 107-108)

In the Maudlin coincise expression: "once what the ontology *is* has been made clear, then (and only then) can one go on to ask what the ontology *does*, how it behaves." (Maudlin 2016, p. 318). To these two steps – the stipulation of the basic constituents and the stipulation of the evolution law for these constituents – there corresponds a difference in the ontological status. The basic constituents are taken to be *primitive*, whereas the laws prescribing the time evolution are *not* as primitive, since they *presuppose* basic constituents in order to operate, i.e. in order to express their lawhood, so to say. In these terms, the monism that the PO approach suggests apparently raises the problem of what the relation between the level of the primitive constituents and the level of the laws constraining them exactly is supposed to be and, consequently, to what extent can we remain neutral on the issue of the ontological status of *laws themselves*[12].

No matter what the answer to this problem may be, however, my focus here is more specific: to what extent can *the Bell notion of beable* live happily within the PO framework? And with reference to which Bell theory of beable – either the first, Bohrian-sounding one or the more mature one – can we answer the above question? In the above presentation of the PO approach by Maudlin, the PO is tightly connected with the theory of beables, since the very task of the primitive ontology approach is presented exactly in terms of the (second Bell formulation of the) notion of beable: the question of how the ontology behaves, namely how laws constrain the basic constituents of the world "is answered by a dynamics: a mathematically precise characterization of how the *beables* change through time." (Maudlin 2016, p. 318, my emphasis). Similarly, Michael Esfeld states that

> the primitive ontology consists in one actual distribution of matter in space at any time (no superpositions), and the elements of the primitive ontology are localized in space–time, being "local beables" in the sense of Bell, that is, something that has a precise localization in space at a given time. (Esfeld 2014, p. 99).

The assumption of a substantial equivalence between the theory of beables and the PO approach, however, has been questioned (Allori 2015, 2021). According to Allori, "some local beables do not make good primitive ontologies" (2021, p. 17), and in order for a PO approach to play suitably its foundational role, some further conditions should be required: not only (i) the specification of a well-defined ontology[13], but also (ii) the possibility for the specific PO formulation to provide *explanations* and (iii) the possibility of defining symmetries for the law(s) constraining the time evolution and behaviour of the basic constituents. A theory of beables in the Bell terms (in the *second* variant) addresses just the task (i), whereas the further conditions (ii) and (iii), in fact, *restrict* the class of the logically possible theories of beables, mainly by imposing structural and nomological constraints on an unqualified set of basic entities:

> Therefore, primitive ontologists are stricter […] in the criteria for a desirable theory: not only do they want a precise ontology […], not only do they want a three-dimensional ontology […], but they also want constructive, dynamical explanations and they want to keep as many symmetries as possible. More explicitly, a *primitive ontology is the simplest local beable that allows for a dynamical, constructive explanation which preserves symmetries*. (Allori 2021, p. 27).

---

[12] Oldofredi 2022 mentions the reductive character of the PO approach, but with reference to the issue of whether there should be one single category of entities *within* the set of the basic constituents. The point I raise here refers instead to the issue whether fundamental reality might admit different sets of 'entities', one including object-like entities and one including law-like entities. But this is a topic for a different paper.
[13] Here I do not touch the relevant issue of the distinction between a *local* PO approach and a possibly *non-local* PO approach, since it is not involved in my main argument: on the point, one can see Allori 2021, sect. 2.

Now I wish to emphasize that the reconstruction of the Bell proposal of a theory of beables in terms of a two-stage process, outlined in the previous sections, can be evaluated on the background of the Allori proposal of characterizing primitive ontologies through a suitable refinement in terms of explanatory and symmetry-related requirements. In other words, my reconstruction provides a historically-based instance of a theory of beables that *cannot* be mapped onto a decently formulated primitive ontology in any sense. According to the early Bohr-inspired theory of beables, put forward by Bell in the first appearance of this notion in the area of the foundations of quantum mechanics, beables are taken to be just "the settings of switches and knobs on experimental equipments, the current in coils, and the readings of instruments". Due to the ambiguity of the apparatus-system cut, implicit in the Bohrian reading of the measurement process considered above – an ambiguity that Bell appears initially to overlook – this early theory of beables is in fact unable to tell, in Bell's terms, what is 'system' and what is 'apparatus': as argued above, this choice of beables may achieve the 'functionalistic' recovery of subjective experience, but fails to play the 'constitutive' role that the recent PO approach assigns to beables. Namely, on the background of the Allori characterization, these beables fail to satisfy even the requirement (i) of representing a well-defined ontology, so that the additional requirements (ii) and (iii) are simply out of the question. So the early, Bohr-inspired version of a theory of beables by Bell turns out to be a factual instance of a theory of (local) beables that *does not* make a good primitive ontology, i.e. a theory which is unable to meet the standards set by the conditions proposed by a robust PO approach.

Finally, I would like to focus on a condition the theory of beables and the PO approach appear to share, namely *theory-relativity*: as stressed by Norsen

> «beable» refers not to what *is* physically real, but to what some candidate theory *posits* as being physically real. Bell writes: «I use the term 'beable' rather than some more committed term like 'being' or 'beer' to recall the essentially tentative nature of any physical theory. Such a theory is at best a candidate for the description of nature." (Norsen 2009, p. 279).

This requirement implies that what has been selected either as beables in the sense of Bell or as as elements of a primitive ontology for a theory $T$ – be they particles, fields, densities of variably defined 'stuff', and so on – can turn out to be not as fundamental in a different theory $T'$ (Oldofredi 2022). In both the theory of beables and the PO approach, therefore, two components appear to coexist. On the one hand, the rejection of a purely instrumental and operational reading of quantum mechanics and the definition of a clear ontology as a necessary step toward the aim of providing secure foundations for the theory. On the other hand, the acknowledgment that the definition of a clear ontology is not a once-and-for-all decision and that there is in fact a pluralism of ontologies compatible with the mathematical structure of quantum mechanics: as a consequence, having a positive attitude toward the PO approach as a fruitful approach to the issue of the foundations of quantum mechanics need not imply the ultimate choice of an ontology at the expense of the others. In this sense, the kind of philosophical stance required by the work within the conceptual environment shared by the theory of beables and the PO approach, an environment sympathetic to the relevance of metaphysics for physics, shows some analogies with a peculiar, meta-theoretical stance on the task of metaphysics, in which "much metaphysical work, especially of the contemporary systematic kind, might best be understood as *model-building*." (Godfrey-Smith 2006, p. 4). If conceived as a model-building activity, metaphysics diverges from science as to its subject matter – the *fundamental features* of the world – but is not so much as to its methodology, since "some theoretical science, though not all, operates as *model-based science*. Model-based science takes an *indirect* approach to representing complex or unknown processes in the real world." (Godfrey-Smith 2006, p. 4, emphasis in the original). As Laurie Paul describes it:

> We can theorize about the world using models, that is, by constructing representations of the world, and metaphysical theorizing is no exception. Scientific theorizing is often understood in terms of the construction of models of the world, and scientific theories about the nature of features of the world may be understood as models of features of the world. Metaphysical theories about the nature of features of the world may also be understood as models of features of the world. Both fields can be understood as relying on modeling to develop and defend theories, and both use a priori reasoning to infer to the best explanation and to choose between empirical equivalents. (Paul 2012, p. 9)

But how should a model be understood in this context? According to Godfrey-Smith,

> a model is an imagined or hypothetical structure that we describe and investigate in the hope of using it to understand some more complex, real-world "target" system or domain. Understanding is achieved via a resemblance relation, that is, some relevant similarity, between the model and the real-world target system (Godfrey-Smith 2006, p. 4).

Taking inspiration from this view, we might conceive the several options available within a beables/PO approach – particles, fields, densities of variably defined 'stuff', and so on – as "models" in the Paul & Godfrey-Smith sense; consistently with the scientific-realistic inspiration of the approach, these models are true *ontological frameworks* that can be supported according to plausible meta-criteria, such as simplicity, intuitivity, continuity with preceding theories in certain respects, and the like[14]. On the other hand, the counterparts of the "real-world, target system or domain" may be conceived as the range of quantum processes and phenomena that fail to be adequately accounted for in the standard formulation of quantum mechanics and that, on the contrary, we expect to receive a more satisfactory description and explanation under one of the selected models. A sort of 'canonical' target system in my analogy is the measurement process in quantum mechanics or, if we want a more restricted form of target system, the uniqueness of the experimental outcome in the situation in which the measured system is prepared in a superposition quantum state. Still according to the analogy, this process or phenomenon can indeed be taken as a target system that needs to be adequately accounted for. According to the usual presentation[15], we assume standard quantum mechanics to describe measurement as a special kind of interaction, such that with the coupling <measured system + measuring apparatus> determines a joint system whose states are supposed to evolve according to the main dynamical law of the theory, i.e. the Schrödinger equation (at least, up to the time of the measurement). Since in a measurement we are supposed to record an outcome for a physical quantity (which is well-defined for the measured system at hand), there are two possible scenarios: (i) if the measured system's state is an eigenstate of the physical quantity to be measured, the state of the joint system will be a state in which the component referring to the measuring apparatus will be unequivocally associated to the reading of (eigen)value of the quantity pertaining the measured system; (ii) if the measured system's state is not an eigenstate of the physical quantity to be measured, the state of the joint system will be a superposition, each component of which will be the product of measured system's state and the measuring apparatus' state, each corresponding to one of the possible (eigen)values for the physical quantity.

---

[14] See again Paul 2012, pp. 12 ff. A further suggestion might be to extend and adapt the considerations put forward by Carnap in his celebrated paper "Empiricism, semantics and ontology" (Carnap 1956) from linguistic to *ontological* frameworks.

Now in the (ii) situation, namely when the measured system is in a superposition before the measurement takes place, the measurement problem amounts exactly to the fact that the following conditions cannot hold together:

**C** – The wave-function associated to the state of a system is a *complete* description of the state itself, namely there can be no finer specification of the properties that the system can exhibit in the event of a measurement;

**L** – The wave-function associated to the state of a system always evolves according to the Schrödinger equation;

**D** – Measurements always provide have determinate outcomes, namely at the end of the measurement process the measuring apparatus is found to be in a state that indicates which among the possible values turns out to be *the* outcome of the process itself.

If we call *objectification* [16] the determinateness at the macroscopic scale of the experimental outcome at the end of the measurement process, objectification appears then to be exactly the sort of target-system for which different ontological frameworks such as Bohmian mechanics – a *particle* framework – or the GRW dynamic reduction approach - a *field* framework – represent different models in the Paul &Godfrey-Smith sense[17], that address in different ways the problem to provide an unambiguous description/explanation for the objectification[18].

---

[15] I refer here to the neat presentation in Maudlin 1995.
[16] Busch, Lahti, Mittestaedt 1991, p. 33.

## 5 Conclusions

John S. Bell introduced the notion of *beable*, as opposed to the standard notion of *observable*, in order to emphasize the need for an unambiguous formulation of quantum mechanics. In the above pages I argued that Bell formulated in fact *two* different theories of beables. The first theory is somehow reminiscent of the Bohr views on quantum mechanics, and is curiously adopted by Bell as a critical tool *against* the Copenhagen interpretation: as I have attempted to show, however, this theory fails to achieve its expected result. The second, more mature theory has been instead among the sources of inspiration of a fruitful interpretive approach to quantum mechanics, called *Primitive Ontology* (PO) approach, that turns out to be inspired to scientific realism. The relationship between the theory of beables and the details of the PO approach is in fact complex: they are not equivalent but share some common conditions. Among these, the *theory-relativity* – namely, some entities may have the beable status in one ontological framework but not in another – has been shown to exhibit some interesting similarities with a view of metaphysics as a model-building activity, a view developed by philosophers who conceive metaphysics and science as enterprises that may diverge on subject matter but possibly converge on methodology of investigation. On the background of the complexity of the relationship between the theory of beables and the variants of the PO approach, finally, the distinction between the two Bell theories of beables allows one to see that the first theory is an early instance of an account of the notion of beable that cannot work as a viable primitive ontological framework.

---

[17] Should we turn to an ontological framework underying the many-worlds interpretation (MWI), I think that among the 'target-systems' we would find not the objectification – a process that, in fact, simply would not occur in the MWI ontological framework – but rather the subjective perception about the determinateness of the outcome at the end of a measurement process, a fact that in MWI is one of the major issues to account for.

[18] Paul 2012 casts his view of metaphysics as a model-building activity within the semantic approach to scientific theories, in which the latter are taken to be suitable classes of models. Should we adopt this approach also in the case of the objectification outlined above, we might define the Bohmian *theory* as the class of all models accounting for objectification in terms of particles in motion, and the GRW *theory* as the class of all models accounting for objectification in terms of fields that spontaneously localize according to specific laws.